# Supportive 5G Infrastructure Policies are Essential for Universal 6G: Assessment using an Open-source Techno-economic Simulation Model utilizing Remote Sensing

**Edward J. Oughton[1] and Ashutosh Jha[2]**
[1]College of Science, George Mason University, Fairfax, VA 22030 USA
[2]S.P. Jain Institute of Management and Research (SPJIMR), Mumbai, 400058, India

Corresponding author: Edward J. Oughton (e-mail: eoughton [at] gmu [dot] edu).

This work was supported by George Mason University and S.P. Jain Institute of Management and Research.

**ABSTRACT** Work has now begun on the sixth generation of cellular technologies (`6G`) and cost-efficient global broadband coverage is already becoming a key pillar. Indeed, we are still far from providing universal and affordable broadband connectivity, despite this being a key part of the Sustainable Development Goals (Target 9.c). Currently, both Mobile Network Operators and governments still lack independent analysis of the strategies that can help achieve this target with the cellular technologies available (4G and 5G). Therefore, this paper undertakes quantitative assessment demonstrating how current 5G policies affect universal broadband, as well as drawing conclusions over how decisions made now affect future evolution to 6G. Using a method based on an open-source techno-economic codebase, combining remote sensing with least-cost network algorithms, performance analytics are provided for different 4G and 5G universal broadband strategies. As an example, the assessment approach is applied to India, the world`s second-largest mobile market and a country with very high spectrum prices. The results demonstrate the trade-offs between technological decisions. This includes demonstrating how important current infrastructure policy is, particularly given fiber backhaul will be essential for delivering 6G quality of service. We find that by *eliminating* the spectrum licensing costs, 100% 5G population coverage can viably be achieved using fiber backhaul. Therefore, supportive infrastructure policies are essential in providing a superior foundation for evolution to future cellular generation, such as 6G.

## I. INTRODUCTION

A flurry of engineering research on 6G is now underway [1]–[7]. Already the provision of global broadband coverage to both unconnected and poorly connected users has been a central development theme [8]–[12]. This topic received less attention than preferred in the previous 5G R&D standardization process. Broadband connectivity is becoming increasingly important to ensure sustainable economic development. There is a particular focus on reducing the digital divide in low- and middle-income countries to support the delivery of the United Nation's Sustainable Development Goals. The global coronavirus pandemic has only increased the political impetus for broadband deployment because it makes digital connectivity even more essential [13], [14].

One of the most cost-effective approaches for delivering broadband over wide geographic areas is via cellular technologies, particularly using 4G, but in the future, this may include 5G too. These cellular technologies are efficient at moving large quantities of data, thus lowering the delivery cost per bit. However, rural connectivity has generally been an afterthought in cellular standardization, meaning the business case for deployment is often weak [15]. Many 6G papers are focusing mainly on urban scenarios, which would lead this generation into the same issues as 5G [16]. Indeed, questions are being asked if 6G needs to play more of a role, whether by new technologies or spectrum management innovation [17]–[21]. Therefore, an emerging aim for 6G is to achieve a dramatic price reduction in cost compared to previous technologies [22], [23]. Our conjecture is that 5G focused too much on providing higher capacity but not enough on reducing cost and providing affordable broadband for the unconnected.

Even with the technologies standardized, the engineering community as well as Mobile Network Operators (MNOs) and governments, still lack effective open-source analytics to help them understand the investment strategies for universal broadband, particularly how these strategies play out in spatio-temporal terms (which is almost always overlooked in both 5G and 6G research) [24], [25]. This provides strong motivation for this paper's content, which aims to consider both the technologies we have available for deployment now (4G and 5G) but approach their evaluation with consideration for a post-5G world ('Next-G'), particularly given the emerging research on 6G technologies. Although the deployment of 6G is still many years away, numerous high-level 6G positioning papers have been published focusing on the qualitative theoretical discussion of 'what should 6G be?' [26]–[35]. We believe we need to start considering the long-term evolution of current technologies to 6G now, but with a greater quantitative focus on cost-effectiveness (with this paper being a demonstrable example).

Despite the grand policy goals for the next decade, we are left with many engineering and economic questions regarding broadband deployment in unconnected locations. When will 5G reach unconnected users? How will decisions we make now prevent further transition to 6G when terabit per second (Tbps) capacity and micro-second (μs) latency are expected? With these issues in mind, the following research contributions for this paper are identified:

1. Assessing how different 4G and 5G strategies quantitatively perform in viably delivering universal broadband coverage.
2. Evaluating the impact that spectrum price changes have on coverage-focused universal broadband strategies.
3. Identifying conclusions to inform current 5G policies and future 6G standardization and deployment.

The remainder of this paper is structured as follows. The next two sections provide an overview of the related literature, followed by an articulation of the generalizable research method in Section IV. The application of the method is presented in Section V, with the results reported in Section VI. A discussion is then undertaken in Section VII which addresses the first two contributions based on the results obtained. The limitations of the assessment are presented in Section VIII. Finally, the third contribution is addressed in Section IX as relevant conclusions are identified.

## II. WHY 5G POLICY MATTERS TO ENGINEERS

In recent years 5G has become wrapped up in an international competition between nations, for example, between the USA, China, South-Korea, Japan, the UK and Europe [36], [37]. There has been a focus on new technological opportunities to provide enhanced capacity and coverage [38]–[44], as well as the cybersecurity issues that could arise [45]–[50].

However, deploying advanced 5G technologies is hitting various economic roadblocks. Firstly, the Average Revenue Per User (ARPU) in mobile markets has either remained static or been in decline in most countries, falling by approximately 1% annually [51]. This is troubling for MNOs who are likely to experience little in the way of new revenue from 5G but are simultaneously being pressured by governments to make large infrastructure investments that deliver on the three main use cases of Enhanced Mobile Broadband (eMBB), Ultra-Reliable Low Latency Communication (uRLLC) and Massive Machine Type Communication (mMTC) [52]. Secondly, the 5G regulatory burden being placed on MNOs is considerable, with significant resources allocated to purchasing spectrum licenses, which could leave little available capital for expansion to less viable locations [53]. These issues do not bode well for deploying 5G to less attractive regions, which could reinforce the digital divide.

Recent literature concerning the deployment of 5G has mainly focused on the policy and economic implications for high-income economies, with only a few examples considering the implications for low- and middle-income countries where most unconnected users reside [54], [55]. Even in leading economies, the policy landscape is still evolving to work out how best to help deliver the potential benefits of 5G, particularly given the embryonic deployment of these technologies. But what has not changed is the desire to extract the maximum amount of spectrum revenue from the sale of new licenses, which to a certain extent is at odds with the policy desire of providing ubiquitous high-capacity broadband connectivity to help spur the digital economy. In summary, there needs to be a much greater quantitative focus on how we will deliver universal broadband at a practical level, including quantification of the ramifications of national policy decisions, for example, on spectrum pricing.

## III. DELIVERING UNIVERSAL BROADBAND

Universal service is a policy that aims to provide all households and businesses with access to a given utility, such as broadband, electricity, or water [56], to be able to reduce access inequality [57]. One of the oldest examples includes universal access to fixed telephone services, which have existed for almost a century [58], [59]. Still, as demand for legacy services has declined, requirements have been adapted to keep up with the digital economy's growth and demand [60]. New universal service policies have also been frequently introduced, particularly when a single previously nationalized service provider is privatized and opened to market forces [61]. In such a case, the policy aim is to ensure that users in locations of market failure, where the cost of supply exceeds the amount that users are willing to pay, do not undergo a loss of service, while simultaneously taking advantage of the benefits of competitive markets in viable locations [62]. Depending on the historical evolution of a telecom market, this can differ by country [63], with some instead favoring the reduction of prices for underserved households [64], [65].

More recently, universal service requirements have been applied to mobile broadband markets via new spectrum licensing regimes. This has enabled the delivery cost to be subjected to market efficiencies via the auction bidding process [66], simultaneously delivering on equity and efficiency objectives [67]. Different designs have been implemented in many countries, each reflecting heterogenous institutional preferences, such as the degree of market involvement and the level of top-down government control [68]–[70]. There are mixed results, however. Although universal broadband aims are admirable, many people are still not connected to a decent service, indicating mixed success in achieving broadband policy objectives.

## IV. OPEN-SOURCE TECHNO-ECONOMIC ASSESSMENT

A generalizable model is now presented, which enables the techno-economic assessment of universal broadband strategies using either 4G or 5G (but could also be adapted in the future to evaluate candidate 6G technology strategies). The fact the code is open-source is desirable because there has been a reproducibility crisis in science over the past decade, giving rise to the need for researchers to share both data and code with other researchers to ensure reproducible results [71], [72]. Thus, an open-source software codebase is developed which enables swift application to any country in the world [73]. The assessment utilizes both simulation techniques and a scenario approach to provide the ability to ask 'what if' questions, which is a common methodological approach for infrastructure assessment [74]–[77], as applied here to a 'hypothetical MNO'. The aim is to use average information to broadly represent market share, spectrum portfolio, and sunk investments in assets to provide a general understanding of different strategic decisions on cellular technologies. This enables a generalizable assessment method to be developed, as visualized in Figure 1. This approach is referred to as a 'generalizable assessment method' because the framework can be more easily applied in other countries thanks to the main model inputs using globally available remote sensing datasets.

A set of scenarios can be used to explore different user capacities. The targets are segmented based on urban, suburban, or rural settlements, reflecting the fact that the engineering requirements and thus the economic costs of delivery are significantly different between these locations. Current universal broadband targets being used by the UN Broadband Commission range from 2 Mbps (enabling most web browsing and email activities) up to 10 Mbps (enabling HD multimedia).

In terms of strategies, there are a wide variety of technologies available for MNOs. Firstly, cellular technologies have proven to be cost-effective in providing wide-area connectivity [38], particularly in locations with no existing fixed broadband infrastructure to upgrade. Either 4G or 5G technologies are the main options currently being considered for broadband connectivity. Secondly, while there are significant choices to make in terms of RAN technologies, the backhaul connection is also an important consideration to provide a cost-effective link from the cell tower to the nearest fiber Point of Presence (PoP) [78]. In many countries, wireless backhaul is still the dominant technology because the costs of deployment are lower than other options.

### A. HIGH-RESOLUTION DEMAND ESTIMATION

A demand assessment framework is developed based on the Mobile Call Termination Market Review 2018-2021 model of Ofcom [79], the UK's telecommunication regulator. The bottom-up Long Run Incremental Cost (LRIC) model used by Ofcom adheres to the International Telecommunication Union's regulatory accounting guidance [80] and is spreadsheet-based. Therefore, the novelty here is the translation of this approach into a spatially explicit representation of demand.

The number of local users for different data services must be estimated, which is a function of the local population, the number of cell phone users, and the types of cell phone users. To obtain the total addressable market for cellular services in the $i$th local statistical area, the population is required ($Population_i$). Using the 1 km² WorldPop population dataset, derived from global satellite imagery, it is possible to extract an estimation of the local population for any location in the world [81]. Via national adoption data, the percentage of cellphone users can then be introduced to obtain an estimate of adoption in the $i$th local statistical area ($CellPen_i$). Additionally, national adoption data on the percentage of smartphone users can also be introduced to provide an estimate of smartphone adoption locally ($SPPen$). Thirdly, the hypothetical MNO only carries traffic for its subscribers. Hence, users are segregated across the available networks uniformly, by dividing the user base by the number of networks in operation ($Networks$). As we aim to deliver 4G and 5G services to smartphone users (as users need this type of device to access them), we thus estimate the number of smartphone users ($SPUsers_i$) in the $i$th local statistical area as in eq. (1).

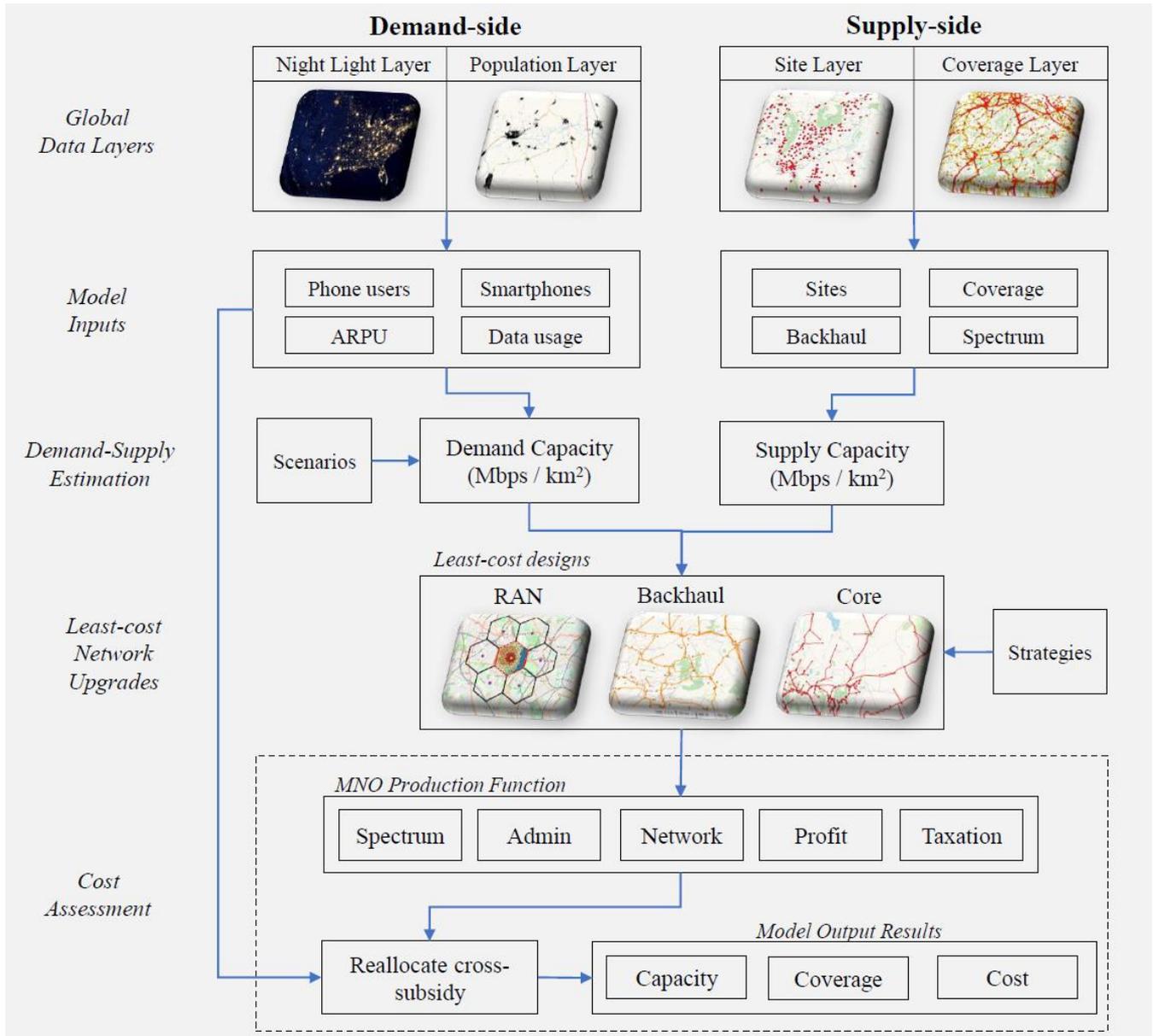

Figure 1 Structure of modeling approach

$$SPUsers_i = \frac{Population_i \cdot \left(\frac{CellPen_i}{100}\right) \cdot \left(\frac{SPPen}{100}\right)}{Networks} \quad (1)$$

This demand equation means that local estimates can be made of cellphone subscribers and smartphone users by network market share, which when aggregated, match the national adoption statistics for the modeled country.

The revenue generated locally ($Revenue_i$) can also be estimated in each local statistical area by allocating consumption tiers to local users based on nightlight luminosity. Using nightlight luminosity remotely-sensed via satellites is an established way to differentiate geographic regions based on the estimated level of development [82].

Hence, this approach can be used to estimate the Average Revenue Per User ($ARPU_c$) for cellular users, broadly segmenting local statistical areas with low luminosity into lower ARPU categories and higher luminosity into higher ARPU categories. The logic is based on local statistical areas with higher socioeconomic status being able to afford to spend more on consuming electricity, which is therefore correlated with being able to spend more on cellular services.

Using the NOAA DMSP-OLS global nightlight layer, luminosity levels are allocated a 'Digital Number' (DN) ranging from 0 to 64 (from no luminosity to very high luminosity) [83]. We allocate local statistical areas above 3 DN into the higher ARPU category, local statistical areas below 1 DN into the lowest APRU category, and local statistical areas falling between into the middle ARPU category.

In eq. (2), we then convert these estimates into the revenue per local statistical area (km²) given consumption of smartphone ($SPUsers_i$) and regular cell phone users ($CellUsers_i$).

$$Revenue_i = \frac{(SPUsers_i \cdot ARPU_c) + (CellUsers_i \cdot ARPU_c)}{Area_i} \quad (2)$$

Future revenue needs to be discounted to the Net Present Value (NPV) over the assessment period to account for the time value of money due to inflation (a monetary phenomenon that occurs from increasing prices over time). The magnitude of this discount rate needs to be based on an expectation of future inflation. For example, the International Monetary Fund (IMF) consumer price projection for India in 2021 is 5% [84].

There also needs to be an estimate of the quantity of user-generated data to design a suitable network to transport this traffic. The estimated level of data traffic ($Traffic_i$) in each local statistical area (km²) is calculated for the given number of smartphone users ($SPUsers_i$) and the scenario defined capacity target for different urban, suburban or rural settlement patterns ($CapacityTarget_s$) using eq. (3).

$$Traffic_i = \frac{((SPUsers_i \cdot CapacityTarget_s)/OBF)}{Area_i} \quad (3)$$

An overbooking factor ($OBF$) is used to reflect the fact that not all users connect to the network simultaneously, as identified in the GSMA 5G Guide [85]. This is similar to the exogenously defined active users parameter elsewhere in the literature [86]. Values used in the literature range from 20-50 [87]–[89], depending on how stringent the network design is for a desired quality of service level.

## B. HIGH-RESOLUTION INFRASTRUCTURE ESTIMATION

Often a geolocated site dataset is not available, only estimates of towers by region, requiring a disaggregation to be carried out (see [90]–[94] for tower counts by country). Therefore, for each statistical unit, data are required for the total population ($Population$), the total number of sites ($Towers$), and the percentage population coverage ($Coverage$). To obtain the number of towers ($Towers_i$) in the $i$th local statistical area, the method reported in eq. (4) allows us to estimate using the population ($Population_i$).

All local statistical areas initially need to be sorted using population density, to allocate towers to the most densely populated areas first, as any rational infrastructure operator would act. Once all towers have been allocated, the remaining local statistical areas without coverage have no towers, reflecting areas of market failure and thus no existing connectivity. This approach ensures that when the disaggregated layer is aggregated, the number of towers matches the national data.

$$Towers_i = Population_i \cdot \frac{Towers}{(Population \cdot (\frac{Coverage}{100}))} \quad (4)$$

The disaggregated site estimates undertaken using eq. (4) are then allocated a technology based on the area coverage by 2G, 3G or 4G technologies using Mobile World Coverage Explorer polygons [95].

As the backhaul technology type for each cell site is not available, we utilize data on the composition of technologies for macro cell sites by region [85], which is 1% fiber, 3% copper, 94% wireless microwave and 2% satellite in South Asia. As we do not have spatial data to estimate backhaul type, a sequential probability can be applied, which allocates the percentage of fiber to sites in the densest local statistical areas and the percentage of satellite to the sites in the least dense locations. Copper and microwave are allocated proportionally to the percentage of sites in the middle of the distribution. Importantly, the backhaul composition allocated in this way ensures aggregated estimates match the data source, avoiding additional modeling uncertainty.

Network maps for telecom operators are digitized and used to establish existing sunk investments in fiber. The structure derived is treated as the network edges and then used to estimate the network nodes. Without data to inform the existing nodes, an estimate is also necessary. Hence, a settlement layer is developed where 1 km² cells above a set threshold are extracted from the raster layer, with spatially proximate cells being summed and those exceeding a specific settlement size being added to the agglomeration layer. Fiber connectivity is then treated as existing at any town with over 10,000 inhabitants within 2 kilometers of a core edge, as a rational infrastructure operator would want to maximize the sale of connectivity services to support the building of a long-distance fiber network. We then also connect any regions without a core node, using a least-cost design. The largest regional settlement is connected to the closest existing core node with a new fiber link. Finally, regional fiber networks are deployed, which connect settlements over 10,000 total inhabitants into an existing core node by building a new fiber link. The least-cost fiber network design consists of a minimum spanning tree estimated using Dijkstra's algorithm, providing a cost heuristic reflecting the actual spatial distribution of assets in a local statistical area. This is superior to the assumptions often used by telecom regulators in spreadsheet-based approaches.

### C. SYSTEM CAPACITY ESTIMATION

The least-cost RAN design consists of two main stages, including using a 3GPP 5G propagation model to obtain the spectral efficiency [96] and then estimating the total channel capacity per spectrum band given a spectrum portfolio.

Firstly, there are three main ways to enhance the capacity of a wireless network, such as increasing the spectral efficiency of the technology in use, adding new spectrum bandwidth, and increasing the spectral reuse by building new cell sites. A generalizable system model is used to estimate the capacity of a cellular network based on using a stochastic geometry approach, which is broadly similar to the open-source Python Simulator for Integrated Modelling of 5G [97], [98].

The mean Network Spectral Efficiency ($\bar{\eta}_{area}^{f}$) (bps/Hz/km²) for a carrier frequency ($f$) in a local statistical area is estimated using the average number of cells per site ($\bar{\eta}_{cells}^{f}$) and the density of co-channel sites ($\rho_{sites}$) utilizing the same spectrum band, as defined in eq. (5).

$$\bar{\eta}_{area}^{f} = \bar{\eta}_{cells}^{f} \cdot \rho_{sites} \quad (5)$$

Hence, for all frequencies in use, the capacity of the local statistical area ($Capacity_{area}$) is estimated via the multiplication of the Network Spectral Efficiency ($\bar{\eta}_{area}^{f}$) by the bandwidth of the carrier frequencies ($BW^{f}$) in use, as in eq. (6).

$$Capacity_{area} = \sum_{f} \bar{\eta}_{area}^{f} \, BW^{f} \quad (6)$$

A radio link budget estimation process is undertaken to estimate the spectral efficiency for three-sectored directional macrocells. Firstly, the received power ($Signal_i$) over a given distance for the $i$th path is estimated, as per eq. (7).

$$Signal_i = TxPower_i + TxGain_i - TxLosses_i \\ - PathLoss_i + RxGain_i \\ - RxLosses_i \quad (7)$$

The constituent components of this approach include the transmitter power ($TxPower_i$), transmitter gain ($TxGain_i$) and total transmitter losses ($TxLosses_i$), producing the Equivalent Isotropically Radiated Power (EIRP). As well as the path loss ($PathLoss_i$), receiver gain ($RxGain_i$) and receiver losses ($RxLosses_i$). The path loss is estimated based on the distance between the transmitter and receiver, using the 3GPP ETSI TR 138 901 (v14) channel model for frequencies ranging from 0.5-100 GHz. A log normal shadow fading distribution is used based on the provided 3GPP parameters [99] for different environments. Building penetration losses are added to the path loss estimate, based on a 50% probability of indoor use. A log normal distribution is also used with a mean of 12 dB and standard deviation of 8 dB based on ITU recommendation M.1225 [100]. Distances within 500 meters are treated as within line-of-sight, whereas distances over are treated as non-line-of-sight. A default transmitter height of 30 meters and a default receiver height of 1.5 meters are used, based on the propagation model guidance.

TABLE 1
SINR TO SPECTRAL EFFICIENCY LOOKUP TABLES

| Channel Quality Indicator (CQI) | SINR (dB) | Spectral Efficiency (Bits per Hertz) | |
|---|---|---|---|
| | | 4G (MIMO 2x2) | 5G (MIMO 4x4) |
| 1 | -6.7 | 0.3 | 0.15 |
| 2 | -4.7 | 0.46 | 1.02 |
| 3 | -2.3 | 0.74 | 2.21 |
| 4 | 0.2 | 1.2 | 3.2 |
| 5 | 2.4 | 1.6 | 4 |
| 6 | 4.3 | 2.2 | 5.41 |
| 7 | 5.9 | 2.8 | 6.2 |
| 8 | 8.1 | 3.8 | 8 |
| 9 | 10.3 | 4.8 | 9.5 |
| 10 | 11.7 | 5.4 | 11 |
| 11 | 14.1 | 6.6 | 14 |
| 12 | 16.3 | 7.8 | 16 |
| 13 | 18.7 | 9 | 19 |
| 14 | 21 | 10.2 | 22 |
| 15 | 22.7 | 11.4 | 25 |

The noise value ($Noise_i$) can be estimated for the $i$th path with eq. (8), using Boltzmann's constant ($k$) (1.38e-23), temperature in Kelvins ($T$) (290 Kelvins = ~16 degrees Celsius), frequency bandwidth ($BW$) (Hz) and the User Equipment (UE) Noise Figure (NF).

$$Noise_i = 10log_{10}\,(k \cdot T \cdot 1000) + NF + \\ 10log_{10}(BW) \quad (8)$$

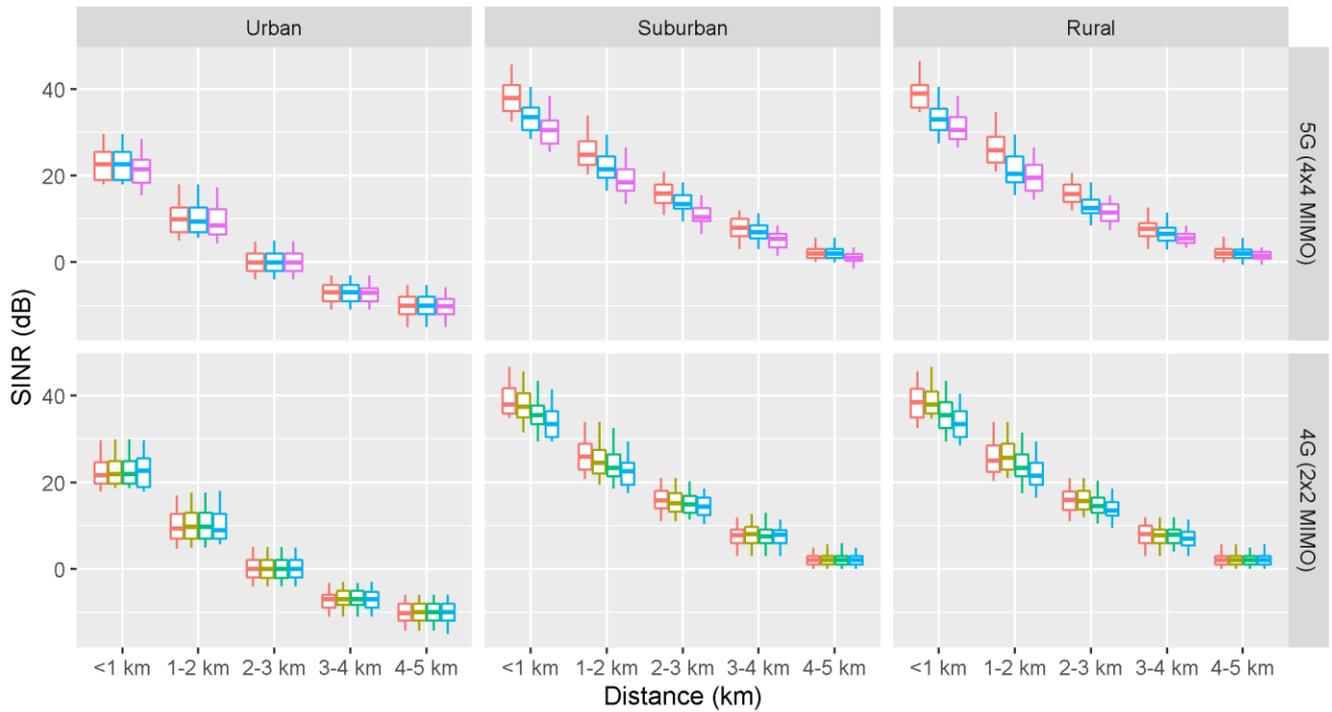
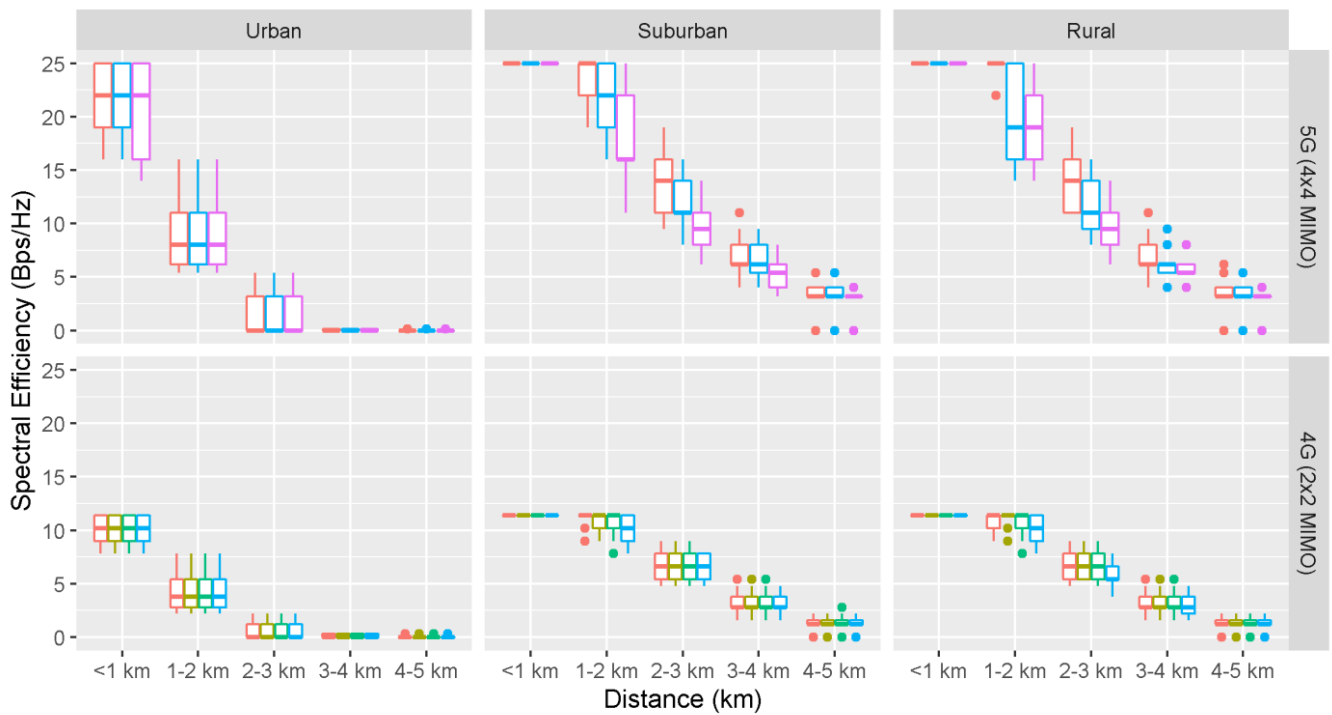

Figure 2 Visualization of SINR and spectral efficiency simulation results

The interference ($I_{i,j}$) for the $i$th path from all neighboring $j$th cells can also be quantified using the received power estimation in eq. (7), enabling the Signal-to-Inference-plus-Noise ratio to be obtained, as per eq. (9).

$$SINR_i = \frac{Signal_i}{\sum_j(Interference_{i,j} + Noise_i)} \quad (9)$$

Once the SINR has been calculated, the corresponding spectral efficiency can be obtained from the lookup tables reported in Table 1, obtained from the literature [96], [99], [101]. Thus, the data transmission rate possible over the wireless link can be estimated.

To estimate the quality of service, the mean capacity provided for the cell (Mbps per km$^2$) is mapped to a particular environment (e.g., urban or rural), antenna type (e.g., 2x2 or 4x4 MIMO), carrier frequency, cellular generation and desired confidence interval. Both the achieved SINR and spectral efficiency values across these different factors are visualized in Figure 2 using box and whisker plots based on the distance from the cell. Initially, using a defined spectrum portfolio, a baseline capacity can be estimated for the current level of infrastructure availability. Then during the modeling process, the same approach can be used to estimate the number of required sites to meet different scenarios of capacity per user, given the number of active users in a local statistical area.

### D. BACKHAUL ESTIMATION

Finally, the backhaul cost to either connect newly deployed cell sites or upgrade the link on existing sites is defined based on the technology strategy being tested and the mean path distance. By accounting for the density of the existing fiber PoPs ($density_i$) in the $i$th region, the mean path distance ($distance_i$) can be estimated ($distance_i = \sqrt{\frac{1}{density}} / 2$). This distance can then be converted to the required fiber investment given the cost per kilometer. For the wireless backhaul, the required investment is also segmented depending on the required distance and the size of the equipment needed. Links under 15 km use a set of small backhaul units, and links over 30 km use a set of large backhaul units, whereas those in between use the medium-sized variant.

### E. COST ESTIMATION

Once a least-cost network has been designed for a particular scenario and strategy, any new greenfield assets or brownfield infrastructure upgrades need to be costed. As there is a time dimension to the assessment study period, all costs are discounted using a 5% discount rate to produce the NPV to the current initial period, which is also informed by IMF consumer price forecasts [84]. The network architecture illustrated in Figure 3 is used to upgrade legacy cellular sites to either of the chosen technologies using the unit cost information reported in Table 2, guided by costs from [89], [102].

TABLE 2
UNIT COSTS

| Component | Cost ($USD) |
|---|---|
| Sector antenna | 1,500 |
| Remote radio unit | 3,500 |
| IO fronthaul | 1,500 |
| Processing | 1,500 |
| IO S1-X2 | 1,500 |
| Control unit | 2,000 |
| Cooling fans | 250 |
| Power supply | 250 |
| Battery power system | 10,000 |
| Base Band Unit Cabinet | 200 |
| Tower | 5,000 |
| Civil materials | 5,000 |
| Transportation | 5,000 |
| Installation | 5,000 |
| Site rental (urban) | 15,000 |
| Site rental (suburban) | 5,000 |
| Site rental (rural) | 1,000 |
| Router | 2,000 |
| Backhaul: Wireless link (small) | 20,000 |
| Backhaul: Wireless link (medium) | 30,000 |
| Backhaul: Wireless link (large) | 60,000 |
| Backhaul: Fiber (m) (urban) | 20 |
| Backhaul: Fiber (m) (suburban) | 10 |
| Backhaul: Fiber (m) (rural) | 5 |
| Regional fiber link (m) | 2 |
| Regional fiber node | 100,000 |
| Core fiber link (m) | 4 |
| Core fiber node | 50,000 |

A literature review is used to evaluate the yielded cost estimates against other cellular deployments for typical three-sector macro cells. The greenfield estimates match an equipment cost of $32k, a site build cost of $20k, and an installation cost of $5k [103]–[110]. Any backhaul or core network upgrades are explicitly modeled based on the distances needing to connect the assets. An annual administration cost is treated as 10% of the capital expenditure, as in prior literature [111].

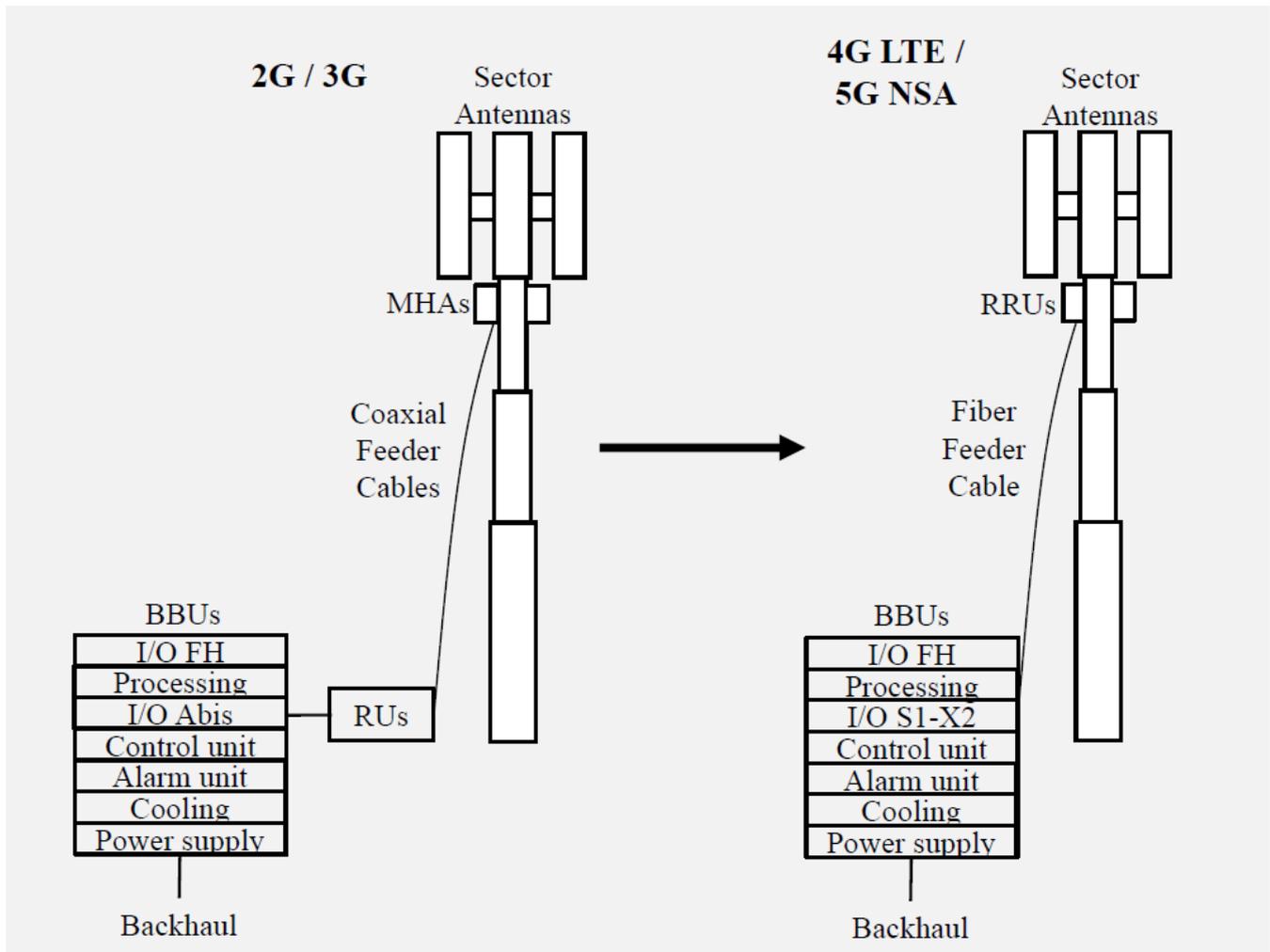

Figure 3 Network architecture for cellular upgrades to 4G and 5G

The cost estimates here do not yet include all the additional administration costs an MNO has to bear, which are added later. For example, these estimates are below the site costs used in other studies, ranging from $100-200k each. Moreover, as the capital needs to be borrowed via money markets, a suitable Weighted Average Cost of Capital (WACC) is applied, reflecting lending risk [112].

Spectrum prices can be developed by taking recent auction results via any available global spectrum database and breaking down each frequency into the US dollar cost per Hertz per member of the population ($/Hz/pop). Such an approach accounts for differences in bandwidth and country population size, which can cause large differences in aggregate spectrum values. Sub-1 GHz bands are treated as 'coverage' spectrum and usually have higher prices due to favorable propagation characteristics. In contrast, frequencies over 1 GHz are treated as 'capacity' spectrum and usually have lower costs due to less favorable propagation characteristics.

Once all these components are combined, the $i$th local statistical area operator cost ($Private\_Cost_i$) is comprised of the investment in the network ($Network_i$), any administration ($Administration_i$), prevailing spectrum prices ($Spectrum_i$), necessary corporation tax ($Tax_i$), and a fair profit margin ($Profit_i$), as illustrated in eq. (10):

$$Private\_Cost_i = Network_i \\ + Administration_i \\ + Spectrum_i + Tax_i \\ + Profit_i \quad (10)$$

To obtain the components of eq. (10), we need to estimate the structure for the network cost, spectrum, taxation, and profit. By taking the sum of the Radio Access Network ($RAN_i$), backhaul ($Backhaul_i$) and core ($Core_i$) in the $i$th local statistical area the Network cost ($Network_i$) can be obtained following eq. (11):

$$Network_i = RAN_i + Backhaul_i + Core_i \quad (11)$$

The admin cost ($Administration_i$) is treated as a percentage of the network and represents the large amount of money that needs to be spent on running an MNO, including on real estate, salaries, vehicle fleets, R&D, etc. This can be up to 30% in high-income economies [113].

Next, to obtain the spectrum cost ($Spectrum_i$) we need to take each of the $f$ frequencies in the $i$th local statistical area and multiply the dollar value per MHz per capita ($Cost\_\$\_MHz\_pop_f$), channel bandwidth ($Bandwidth_f$) and population ($Population_i$), as per eq. (12):

$$Spectrum_i = \sum_f Cost\_\$\_MHz\_pop_f \cdot Bandwidth_f \cdot Population_i \quad (12)$$

For the taxation rate ($Tax\_Rate$) in the $i$th local statistical area, the total tax due ($Tax_i$) can be calculated given the sum of the network cost ($Network_i$) as detailed in eq. (13):

$$Tax_i = Network_i \cdot \left(\frac{Tax\_Rate}{100}\right) \quad (13)$$

As an MNO takes a risk in a private market, there needs to be a fair return for any 4G or 5G infrastructure provision. Therefore, in the $i$th local statistical area, a profit margin ($Profit_i$) is added for all investments (in addition to the WACC risk premium), as stated in eq. (14):

$$Profit_i = (Network_i + Spectrum_i + Tax_i) \cdot \left(\frac{Profit\_Margin}{100}\right) \quad (14)$$

An important part of the model is that excess profits (e.g. >10%) are reallocated via user cross-subsidization to unviable locations to explore how the total revenue in the market could encourage infrastructure rollout in harder-to-reach places. Without such a mechanism, the only viable locations would be dense urban and suburban settlements, and therefore there would not be any further upgrade to other locations (which does not necessarily match reality). After accounting for any reallocated capital via user cross-subsidization, any shortfall in connecting unviable locations would consequently require a state subsidy.

## V. APPLICATION

An assessment period of 2020-2030 is used to capture cellular deployment over the next decade focusing on testing either 4G or 5G Non-Standalone (NSA) strategies. India is used as an example as the country fits with the key trends already identified as affecting the deployment of 5G.

Firstly, India's ARPU has been on a constant decline in recent years, resulting in plummeting revenues for various incumbent MNOs [114]. Amidst such a scenario, there are widespread apprehensions concerning the financial feasibility of deploying 5G networks and provisioning 5G services in the country. Secondly, India is regarded as having some of the highest spectrum prices globally, which raises issues around how aggressive the reserve price may be for 5G bands. Additionally, India has a well-known issue with cellular backhaul availability [115], [116].

TABLE 3
ARPU CONSUMPTION TIERS

| Region name | Code | ARPU Tier ($) | | | Spectrum cost (<1 GHz) ($/Hz/pop) | Spectrum cost (>1 GHz) ($/Hz/pop) |
|---|---|---|---|---|---|---|
| | | Low | Medium | High | | |
| Andhra Pradesh | AP | $0.6 | $1.2 | $1.9 | $2.22 | $0.54 |
| Assam | AS | $0.5 | $1.0 | $1.6 | $0.70 | $0.13 |
| Bihar | BR | $0.4 | $0.8 | $1.2 | $0.19 | $0.05 |
| Delhi | DL | $0.5 | $1.0 | $1.5 | $10.18 | $3.04 |
| Gujarat | GJ | $0.5 | $1 | $1.6 | $1.11 | $0.32 |
| Haryana | HP | $2 | $3 | $6 | $0.89 | $0.25 |
| Himachal Pradesh | HR | $0.4 | $0.8 | $1.1 | $0.67 | $0.28 |
| Jammu & Kashmir | JK | $0.5 | $1 | $1.5 | $0.59 | $0.13 |
| Karnataka | KA | $0.6 | $1.2 | $1.8 | $1.19 | $0.46 |
| Kerala | KL | $0.6 | $1.3 | $1.9 | $1.2 | $0.38 |
| Kolkata | KO | $0.4 | $0.9 | $1.3 | $11.76 | $3.09 |
| Madhya Pradesh | MH | $0.6 | $1.1 | $1.7 | $1.27 | $0.29 |
| Maharashtra | MP | $0.5 | $1.0 | $1.4 | $0.71 | $0.13 |
| Mumbai | MU | $0.6 | $1.2 | $1.9 | $7.39 | $2.29 |
| North-East | NE | $0.5 | $1.1 | $1.6 | $0.50 | $0.09 |
| Orissa | OR | $0.5 | $0.9 | $1.4 | $0.34 | $0.08 |
| Punjab | PB | $0.5 | $1.1 | $1.6 | $1.07 | $0.42 |
| Rajasthan | RJ | $0.5 | $1.1 | $0.5 | $0.58 | $0.23 |
| Tamil Nadu | TN | $0.6 | $1.3 | $1.9 | $1.22 | $0.89 |
| Uttar Pradesh (East) | UE | $0.4 | $0.7 | $1.1 | $0.24 | $0.01 |
| Uttar Pradesh (West) | UW | $0.4 | $0.8 | $1.2 | $3.92 | $1.43 |
| West Bengal | WB | $0.5 | $0.9 | $1.4 | $0.21 | $0.05 |

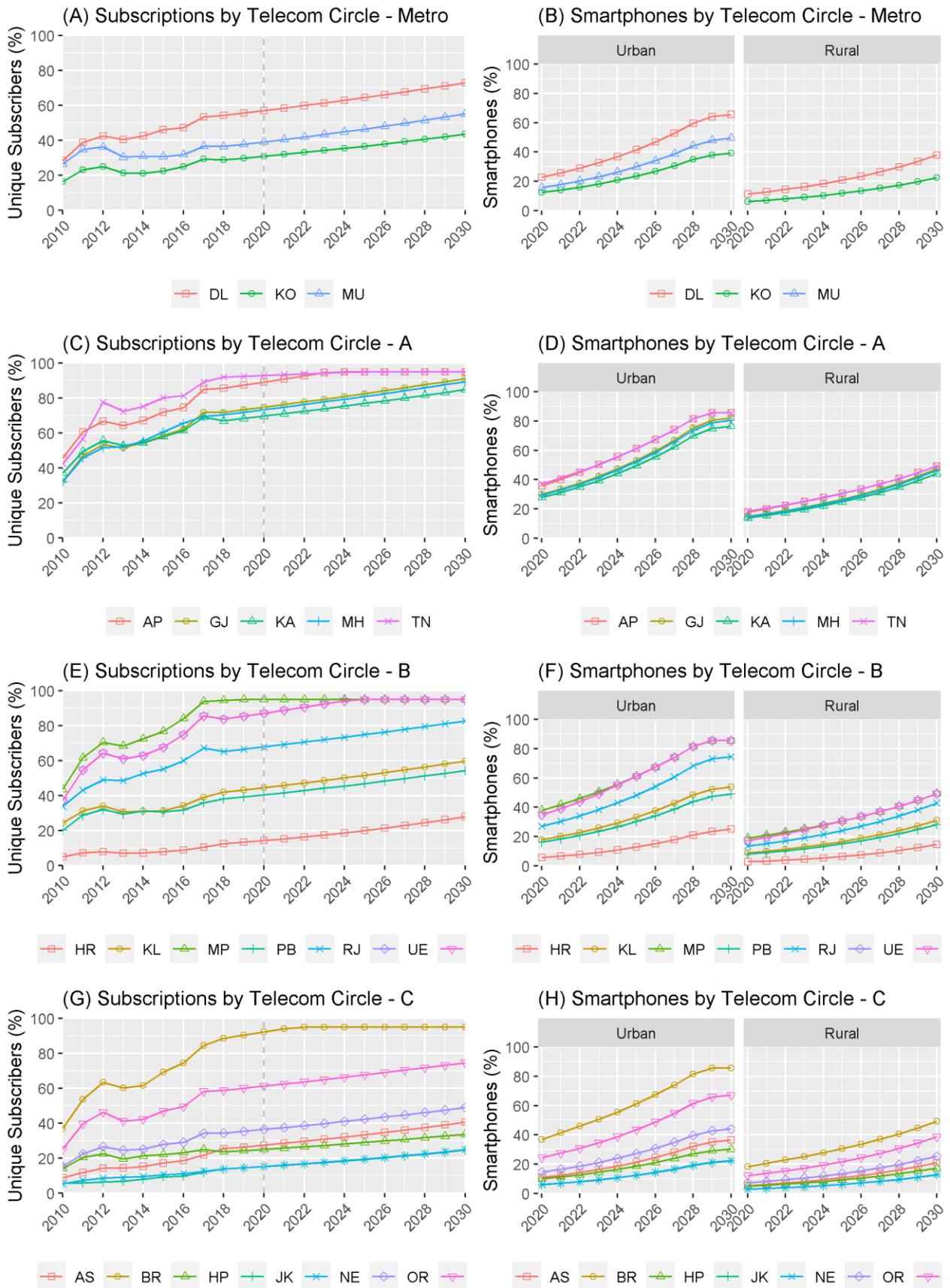

Figure 4 Subscriber and smartphone forecasts

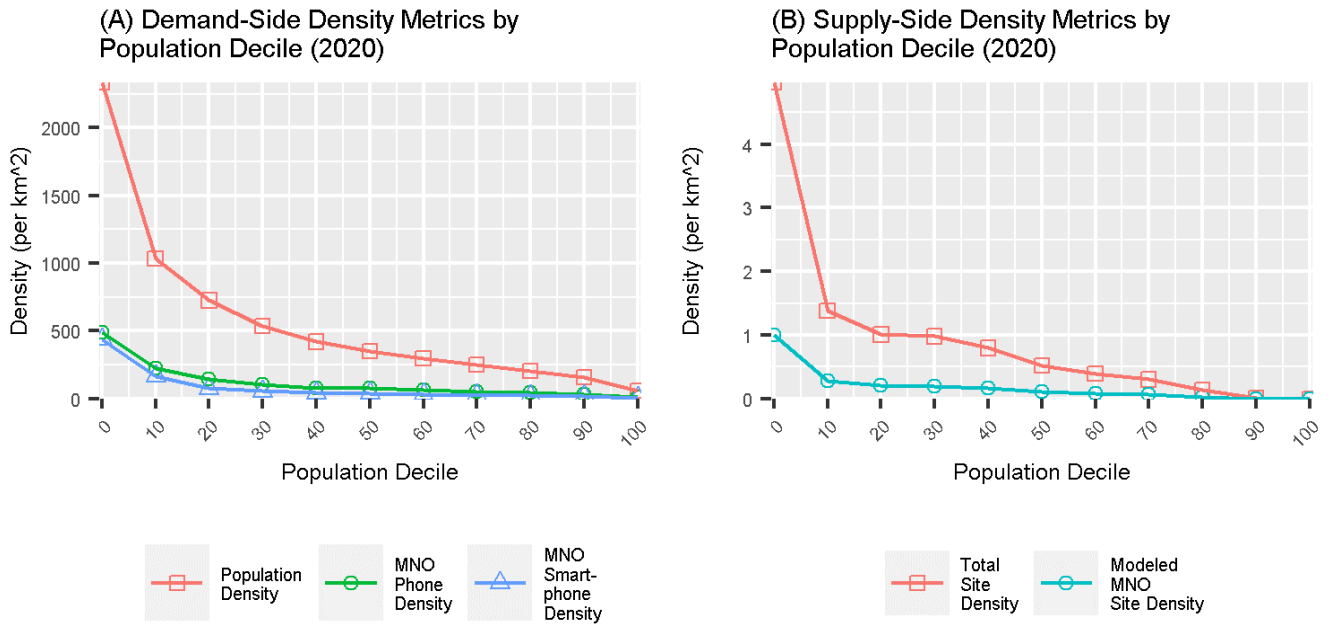

Figure 5 Demand and supply density metrics for the year 2020

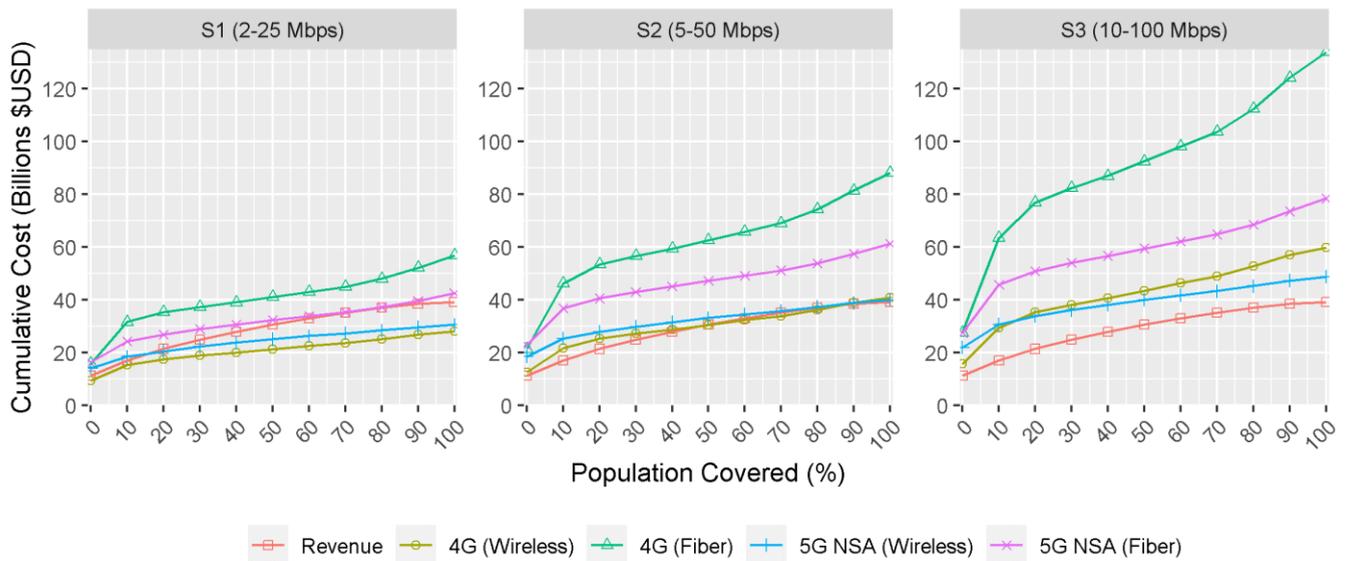

Figure 6: Scenario viability by technology strategy

India is divided into twenty-two wide-area telecom zones, referred to as 'telecom circles', and each comprise of many individual local statistical areas as defined in Section IV. Each telecom circle has a different potential for financial returns and, therefore, different spectrum prices. This creates a considerable administrative burden on an MNO to obtain licenses. In India, researchers have already been evaluating the feasibility of 5G deployment across a wide range of technologies [117]–[130]. With the existing level of capacity between 3-10 Mbps per user, there is considerable scope for improvement, although we should recognize that these estimates are based on crowdsourced data from mainly urban users, so the situation is likely much worse in rural locations [131].

Scenario 1 focuses on a basic set of targets for urban, suburban and rural locations consisting of 25, 10 and 2 Mbps, respectively. Secondly, in Scenario 2, an intermediate set of targets for urban, suburban and rural locations focus on delivering 50, 20 and 5 Mbps, respectively. Finally, in Scenario 3, a highly ambitious set of capacities for urban, suburban and rural locations aim to deliver 100, 30 and 10 Mbps, respectively. The scenarios selected represent a broad range of options to provide insight into how the delivered capacity affects cost, and therefore the deployment of universal broadband using either 4G or 5G across different urban-rural settlement patterns in India.

TABLE 4
SIMULATION PARAMETERS

| Simulation Parameter | Simulation Value | Unit |
|---|---|---|
| Spectrum bands | 850, 1800, 2300, 700, 3500 | MHz |
| Spectrum bandwidth (respectively) | 2.5, 2.5, 15, 5, 50 | MHz |
| Duplex format (respectively) | FDD, FDD, FDD, FDD, TDD | - |
| Inter Site Distance (ISD) | 0.4-40 | km |
| Overbooking factor | 20 | - |
| Tx power | 40 | dBm |
| Tx gain | 16 | dBi |
| Tx losses | 1 | dB |
| Tx antenna type | Directional | - |
| Tx sectors | 3 | Sectors |
| Tx height | 30 | Meters |
| Rx gain | 0 | dB |
| Rx losses | 4 | dB |
| Rx Height | 1.5 | Meters |
| Noise figure | 1.5 | dB |
| Propagation model | ETSI TR 138 901 (v14) (0.5-100 GHz) | - |
| Shadow fading distribution type | Log-Normal | - |
| Shadow fading distribution parameters | $(\mu, \sigma) = (0, \sigma)$ | dB |
| Building penetration loss distribution Type | Log-Normal | - |
| Building penetration loss distribution parameters | $(\mu, \sigma) = (12, 8)$ | dB |
| Frequency reuse factor | 1 | Factor |
| Indoor probability | 50 | % |
| Line of sight | <500 | Meters |
| Transmission method | 4G: MIMO 2x2, 5G: MIMO 4x4 | - |
| TDD DL:UL | 4:1 | - |
| Network traffic load | 50 | Percent |

The telecom circles are listed by name and abbreviation code in Table 3, along with the ARPU consumption tiers per user in each local statistical area. The demand forecasts developed can be viewed in Figure 4 for all regions assessed. The forecasts visualize both the number of unique mobile subscribers and the adoption of smartphones. For the cellular penetration rate, the number of unique subscribers is obtained from the historical data (2010-2020) and used for forecasting over the study period to 2030 [132]. Historical data is not available for smartphone penetration; therefore, a set of consistent growth rates are used to forecast smartphone penetration across both urban and rural regions. In Figure 5, both the demand and supply metrics are presented nationally by decile for India, for both the total market and a single modeled MNO with a 25% market share.

In developing the settlement layer, most telecom circles use a cell threshold of 500 inhabitants km$^2$ with a settlement threshold of 1000 total inhabitants. The exceptions include Mumbai, Delhi, and Himachal Pradesh, which use a cell threshold of 1000 inhabitants km$^2$ and a settlement threshold of 20,000 total inhabitants. The resulting points layer of all settlements is used to develop the least-cost network routing structure. To incorporate both the existing as well as the planned fiber network across the settlements, the geospatial data for the Indian railway network is used, since fiber deployments are laid along the railway lines [133]. If settlements are within a 5 km buffer of the railway line they are treated as having fiber connectivity because the rational aim of deploying the network is to maximize access to as many settlements as possible.

For the supply assessment, the simulation parameters reported in Table 4 are used to undertake the system capacity estimation process, in combination with the generalizable model already presented in Section IV.

An average MNO spectrum portfolio for India is identified, which includes deploying 4G in Frequency Division Duplexing (FDD) using 850 MHz (MIMO 2x2) with 2x2.25 MHz of bandwidth for paired channels (except in Tamil Nadu where 2x1.25 MHz is used). Additionally, 1800 MHz is available with 2x2.5 MHz bandwidth and 2300 MHz with 2x15 MHz bandwidth, both using FDD. For 5G, 700 MHz is the main low band frequency using 2x5 MHz bandwidth for paired channels in FDD (MIMO 4x4). In contrast, 5G can also take advantage of Time Division Duplexing (TDD) spectrum at 3.5 GHz (MIMO 4x4) with a single 1x50 MHz bandwidth channel, with a 4:1 ratio between the downlink to uplink, given the download capacity is the bottleneck in cellular systems.

In terms of other parameters, the MNO administration cost is treated as 20% of the network and the corporation tax rate is treated as 22% of profit, as is the baseline rate in India. The prevailing Weighted Average Cost of Capital (WACC) for India is 10% [112]. Having detailed how the generalizable model is adapted for India's case study example, the results will now be reported.

## VI. Results

The viability of 4G and 5G technologies in delivering universal broadband over the study period are visualized in Figure 6 for the different scenarios and strategies tested. The cumulative cost is used to demonstrate the required investment needed to provide coverage up to each population decile (with deciles sorted based on population density).

Across the scenarios tested, the results demonstrate that the capacity per user is well correlated with the cost of provision, given the required investment increases significantly as the scenarios become more ambitious. Indeed, as the number of required cell sites increases to serve higher demand, this has a major impact on the cost of building fiber connections, with both 4G and 5G fiber-based strategies being the most expensive options. When interpreting the performance of the different strategies tested, the cumulative cost should be compared relative to the cumulative revenue as this demonstrates the level of viability present. In Scenario 1, we can see that both 4G and 5G, both using wireless backhaul, are viable to service 100% of the population, thus delivering universal broadband. In contrast, fiber strategies can only viably serve up to ~70% of the population in the best case.

In Scenario 2, both 4G and 5G NSA using a wireless backhaul can viably provide universal coverage of 100% of the population. This is due to the existing advantage that 4G has in baseline availability, in that there are already a substantial number of sites with this technology in use. In contrast, while 5G is more spectrally efficient, all sites need to be upgraded with this new technology. Finally, in Scenario 3, when trying to deliver up to 100 Mbps per user, all strategies are unviable as this target is too ambitious given the potential APRU.

However, the cost composition of the required investment is different depending on the deployment context, as demonstrated in Figure 7 for each scenario and strategy. There are two main differences visible. Firstly, the proportion that the backhaul cost contributes to the overall investment composition is high in both the most populated deciles and the least populated deciles. In the former, this is the result of needing lots of sites. Whereas in the latter, this is the result of the backhaul needing to traverse a longer distance to the closest fiber PoP. Secondly, the proportion that the spectrum cost contributes varies. In more populated locations, there is a much higher contribution to the cost of the overall spectrum (because of the greater population), whereas, in the final less populated deciles (where there are fewer people), the contribution to the overall spectrum cost is much lower. These two factors lead to an observable pattern across the scenarios and strategies tested. The aggregate cost per decile is generally higher in both the most and least populated locations, whereas the aggregate cost is lower in the middle deciles.

Aggregate costs overlook the number of users served per decile. Therefore, in Figure 8, the required investment is broken down per user. Again, the results are reported by the cost type for each decile across the different scenarios and strategies. There is a strong relationship across the distribution, whereby the cost per user is lower in the first population deciles, where the population density is highest. The cost per user then increases in tandem with the decrease in population density. In Figure 8, it is also useful to view the required cost per user by decile for the study period because this is a much more meaningful number, given monthly and annual ARPU is generally well understood because many people have cellular subscriptions.

Even with 4G using a wireless backhaul, we can see in Figure 8 that $424-883 per user in the most rural decile is going to be challenging (top row), and thankfully the comparative cost for 5G NSA with a wireless backhaul is lower at $299-470 across the scenarios (third row). 5G is cheaper thanks to the use of higher-order MIMO (4x4), enabling the capacity targets to be met using fewer sites compared to 4G (2x2), thereby reducing the required investment cost. Both RAN technologies using fiber are far too expensive for the hardest-to-reach locations, with the cost ranging from $1461-3059 for 4G and $956-1551 for 5G NSA (second and fourth rows respectively in Figure 8).

With spectrum playing a large part in the cost composition of the cheapest technology options, it is worth investigating the impact of changes in spectrum prices on the viability of deployment. This is undertaken in Figure 9 using sensitivity analysis, where a parameter sweep is undertaken of the spectrum cost, to assess how universal broadband viability changes under different cost structures.

Lowering spectrum fees means that MNOs have more available capital to invest in less viable locations, therefore boosting coverage. Such a policy decision would need to be used in tandem with a universal service obligation to ensure the change in the MNO cost structure leads to enhanced infrastructure build-out in harder-to-reach areas. Such obligations could be included in a spectrum license auction, with a proportion of the license fee returned for each percentage of additional coverage an MNO achieves.

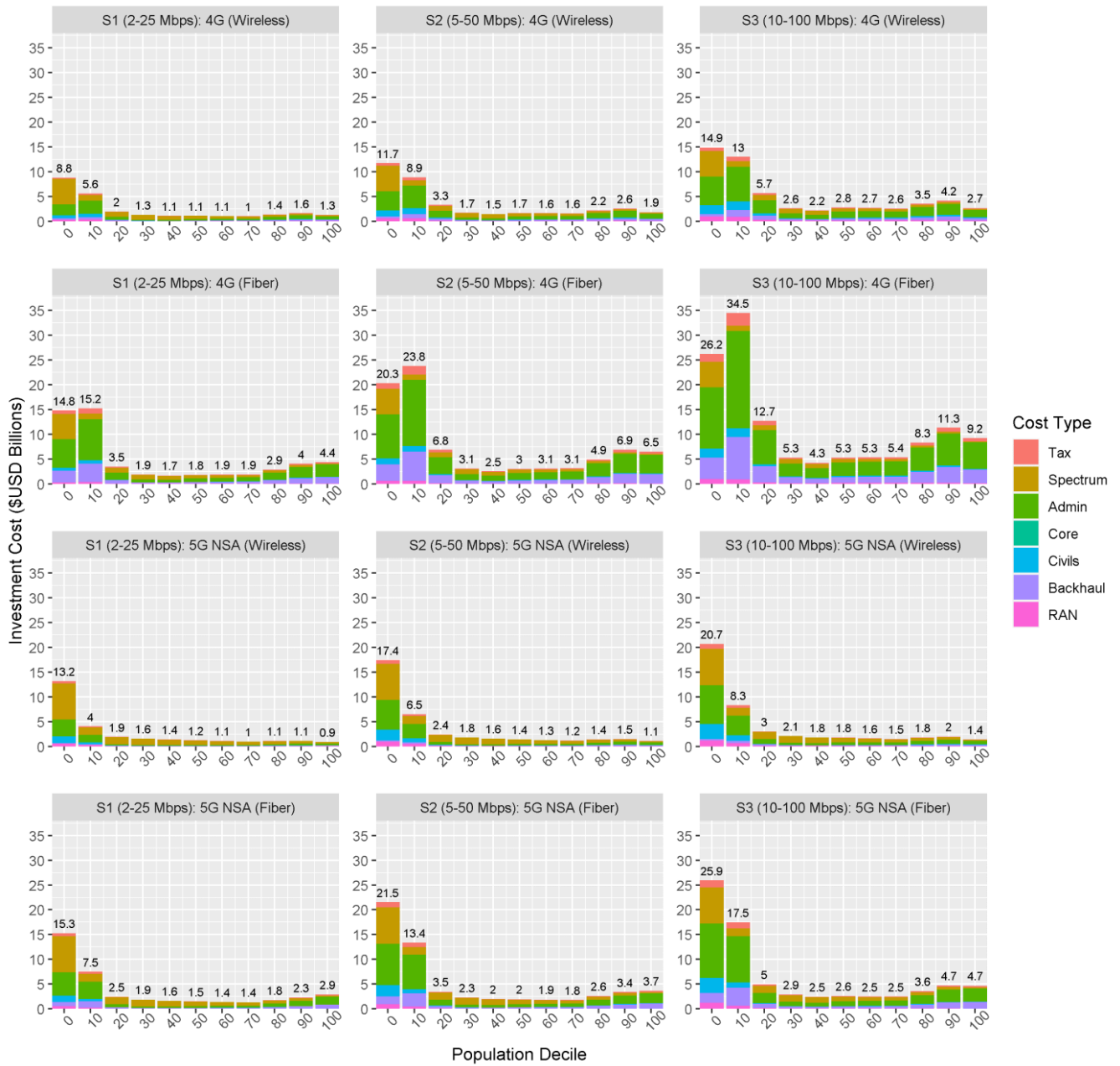

Figure 7 Required investment by population decile for each scenario, strategy, and cost type

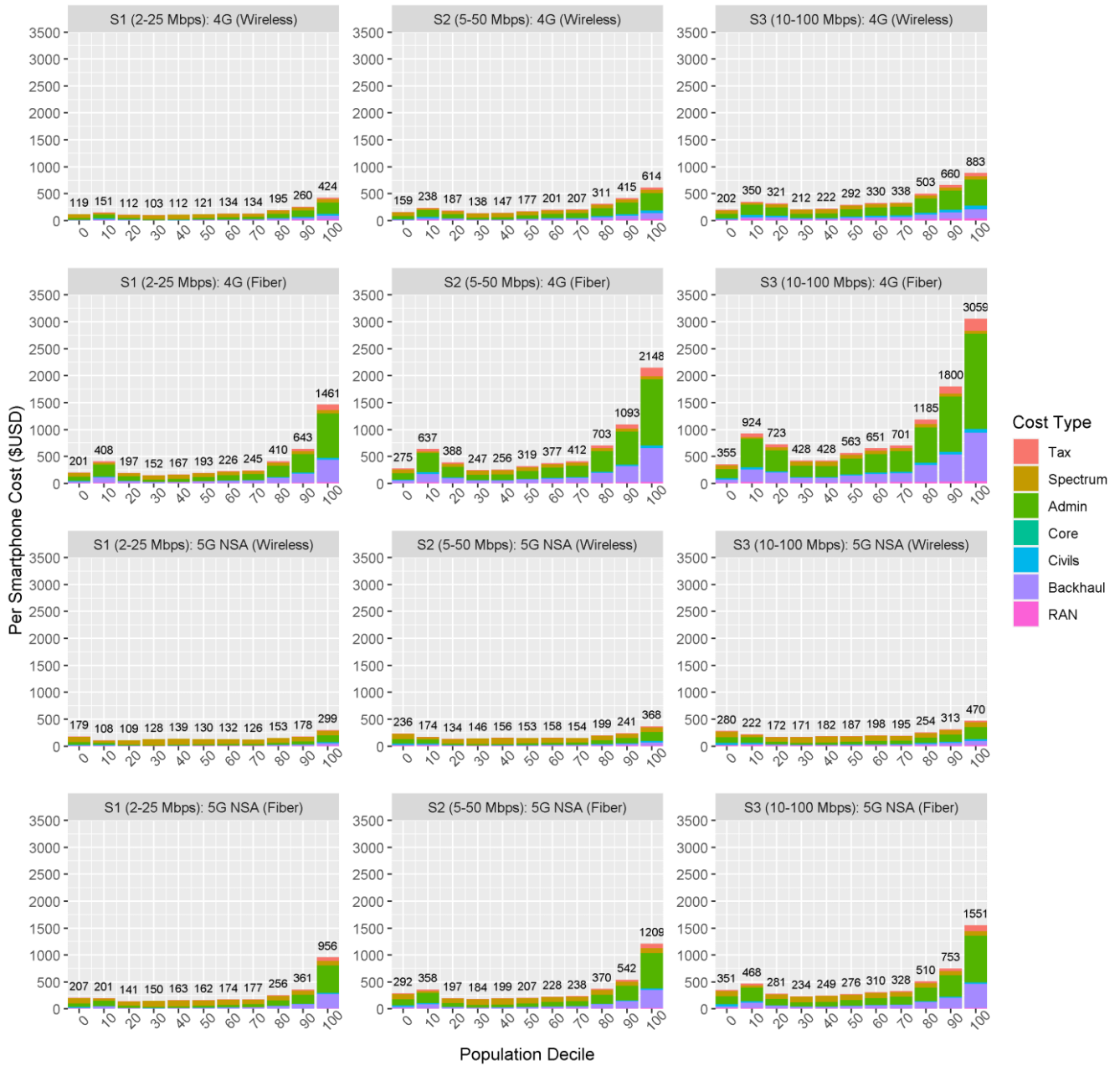

Figure 8 Per user cost by population decile for each scenario, strategy, and cost type

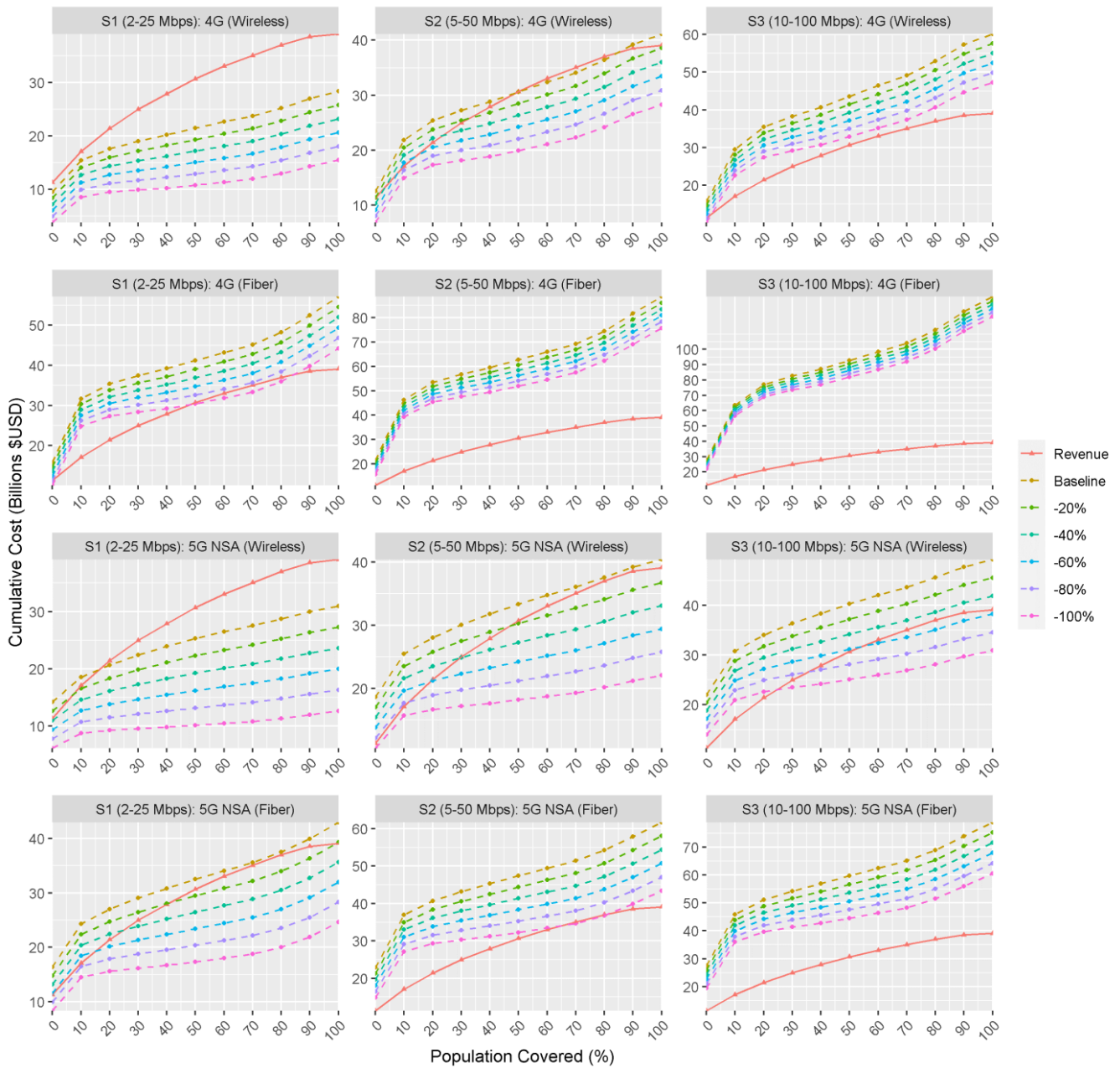

Figure 9 The impact of spectrum costs

In Figure 9, the cumulative revenue across population deciles is plotted against the baseline, as well as different decreases in spectrum prices to evaluate the sensitivity of this cost input. The aim is to evaluate the impact of spectrum price reductions as they filter through into the cumulative cost of deployment against the point at which the cost curve crosses the cumulative revenue. If a particular decile has a revenue curve above the cost curve, the scenario and strategy are viable. In contrast, if the cost is above the revenue, then the scenario and strategy are unviable.

Viability varies across the different scenarios and strategies in Figure 9. With lower capacity per user, such as in Scenario 1, most strategies are either fully viable or close to fully viable with the baseline spectrum price, except for 4G with a fiber backhaul. However, delivering a minimum speed of 25 Mbps in urban and 2 Mbps in rural locations may be perceived as not ambitious enough.

Thus, in Scenario 2 in Figure 9, the available capacity is an improvement, but viability already becomes difficult without resulting to using either wireless backhaul or reducing the spectrum price. For example, 5G NSA with a fiber backhaul is unviable in the baseline, but if spectrum prices were eliminated altogether, it would be possible to viably reach 100% population coverage (although, this may not be politically a feasible option and would only be plausible if universal service obligations were introduced to guarantee delivery). With the most ambitious target in Figure 9, Scenario 3, all strategies are unviable in the baseline. Even with a drastic reduction in spectrum prices, fiber backhaul options are still unviable in all circumstances.

There are important results to take note of in Scenario 3, however. Changes to spectrum costs would not be enough to alter 4G's viability level, but a 60% reduction for 5G NSA using a wireless backhaul would enable coverage to reach 100% of the population.

## VII. DISCUSSION

The assessment presented in the analytical part of this paper used an **open-source modeling codebase** [73] to quantitatively evaluate a range of 4G and 5G universal broadband strategies. A combination of remote sensing and infrastructure simulation techniques was combined to provide insight into the capacity, coverage, and cost of both 4G and 5G infrastructure strategies. The results provide insight into the viability of different strategies, depending on heterogenous per user capacity scenarios, including providing the required investment on a per user basis. Finally, a sensitivity analysis was performed to quantify the impact that governmental spectrum pricing regimes have on the economics of universal broadband connectivity, with ramifications for both short-term deployment and long-term evolution to 6G. This section now discusses their ramifications regarding the first two research contributions articulated in the introductory section of this paper. The first research contribution identified was as follows:

*Assessing how different 4G and 5G strategies quantitatively perform in viably delivering universal broadband coverage*

In terms of the performance of the strategies across the scenarios, the required investment for universal broadband increased as the ambition of the user capacity scenario grew. Generally, the fiber backhaul strategies were much more expensive, supporting the idea that wireless backhaul will remain a key approach for delivering 4G and 5G universal broadband over the coming decade for hard-to-serve locations, should there be no changes in the fiscal landscape. For example, in Figure 6, 100% of the population could viably be served in Scenario 1 (2-25 Mbps) using both 4G and 5G with wireless backhaul, whereas fiber strategies were far less viable. Moreover, total population coverage could be achieved in Scenario 2 (5-50 Mbps) for both 4G and 5G using a wireless backhaul. However, in all circumstances Scenario 3 (10-100 Mbps) was unviable regardless of the strategy, as this target is too ambitious given the potential APRU, which can be very low for rural locations.

The aggregate cost across the deciles modeled exhibited a U-shape pattern. Hence, there was a much higher aggregate cost in both the most and least populated locations but a considerably lower aggregate cost in the middle deciles where the population density is much more amenable to deploying low-cost 4G and 5G broadband connectivity. When considering the required investment per user, there was a strong dynamic where the cost per user was lower in the deciles with the highest population densities, but as the population density decreased, the cost per user inversely increased. This results from scale economies and the need to split the fixed costs in cellular deployment over the local users accessing specific infrastructure connections. This is not unique to cellular and is exhibited in all infrastructure networks, such as transportation, energy, and water.

To provide universal broadband connectivity, we know the most considerable challenge will be in serving the hardest-to-reach locations with the lowest population density. The results show that the costs differ in serving the final population decile depending on the technology deployed. For example, in Figure 8 with 4G using a wireless backhaul, the cost per user in the most rural decile was between $424-883 across the different scenarios. Given how low incomes can be in rural locations, this is by no means an easy target to reach using market methods alone, and state subsidies may be required to provide support for unviable locations. Fortunately, deploying 5G NSA with a wireless backhaul is the cheapest option in these situations, with the cost per user ranging between $299-470 across the scenarios (Figure 8). This compared with much larger per user costs using fiber, where the investment would need to range from $1461-3059 for 4G and $956-1551 for 5G NSA across the scenarios tested (Figure 8). However, the caveat to any 5G strategy would be whether the local population had 5G-enabled handsets to take advantage of the available capacity.

Having discussed the first research contribution, the second will now be evaluated, which was as follows:

*Evaluating the impact that spectrum price changes have on coverage-focused universal broadband strategies.*

Governments have many instruments at their disposal to help reduce the costs of broadband deployment in the hope of achieving universal coverage. High spectrum prices are a well-known issue, particularly for India, the example country assessed here. Therefore, the use of sensitivity analysis for this model parameter in Figure 9 helps provide insight into the ramifications of potential policy changes. As the least ambitious scenario (2-25 Mbps) was either viable or close to viable for most 4G and 5G strategies, there is less relevance here in exploring spectrum price changes, especially as policy ambitions might be aiming higher than the user capacities targeted in this option. However, in Scenario 2 (5-50 Mbps), while 4G and 5G using wireless backhaul was viable for providing universal broadband, there were other interesting results. 4G with fiber was not viable, even with reduced spectrum costs, but 5G NSA with fiber could be plausibly delivered universally if the spectrum cost were *eliminated*. This would obviously take significant political will to make such a bold move and would require affiliated coverage obligations to ensure MNOs deliver the necessary infrastructure but could provide a significant improvement for the availability of broadband connectivity, and also provide a fantastic starting point for evolving to 6G, where fiber backhaul is almost certainly going to be required. Finally, Scenario 3 (10-100 Mbps) provides much more admirable per user capacity. Therefore, it is attractive that only a 60% spectrum price reduction would viably enable 5G NSA using wireless backhaul to provide universal broadband to 100% of the population, under the engineering and economic conditions assessed here.

Having discussed the ramifications of the results for the 4G and 5G universal broadband assessment undertaken, the conclusion will now consider the broader implications, particularly with reference over the long term to universal 6G.

## VIII. LIMITATIONS

Although the method outlined provides an important contribution to the literature, there are limitations that need to be discussed. For example, in the assessment of any cellular network at the national level, simplifications are made. In this analysis, the data-focused assessment excludes the small amount of traffic <10% generated by legacy networks such as 2G or 3G, in preference of assessing current 4G and future 5G traffic, for example in India, meaning the overall traffic may be underestimated. This issue is likely to diminish over time now that legacy networks are being switched off and having spectrum re-farmed to more spectrally efficient technologies (e.g. 4G).

As with any empirical assessment of a telecom market, there are missing data, meaning certain parts of the model require improved estimation. A good example is regarding the way local cellphone and smartphone adoption is estimated in the absence of actual local adoption data. Future research may want to explore techniques, such as integration, to reduce uncertainty in estimating these local model inputs.

Generally, the benefit of undertaking national assessments openly, as is done here, is that future analyses may benefit from government data support, should there be an interest to help rerun the evaluation with the type of market information telecommunication regulators hold. By providing the codebase used here open source, there is hope that other researchers will access the code, explore model limitations and contribute improvements to the approach developed here.

## IX. CONCLUSION

Can conclusions be developed to inform current 5G policy and future 6G standardization and deployment? For example, what do these results mean for universal broadband policy? Are there implications for the 6G R&D effort? Indeed, which issues should engineers researching 6G technologies be cognizant of to achieve our shared aim of universal broadband connectivity? These important questions will now be answered by drawing relevant conclusions, helping to answer the third research contribution articulated in the introduction of the paper.

**The technology choices currently being made have significant long-term trade-offs**. While this may sound platitudinous, this analysis demonstrates that MNOs and governments need to be aware of how backhaul decisions will play out over the next decade and beyond. For example, wireless backhaul methods are clearly the winner in helping to achieve expedited cost-efficient deployment of broadband connectivity in hard-to-reach rural and remote locations. However, if we work from the assumption that fiber is almost certainly going to be required to deploy high-quality broadband connectivity, for example via universal 6G, governments need to be aware that it may make more economic sense to deploy fiber now rather than wireless. Obviously, this takes resources but as the analysis in this assessment has shown, the spectrum revenues extracted from the telecom industry are significant and changes to this framework would enable greater fiber penetration to help deliver broadband connectivity. For example, universal 5G using fiber backhaul could be achieved by eliminating the spectrum cost, enabling this capital to be reallocated to fiber infrastructure investment. While this is a politically sensitive question (as spectrum revenues are alluring for governments), the real issue is the potential benefits gained from providing enhanced broadband connectivity. Indeed, if they outweigh the revenues generated via spectrum licensing then they may warrant a re-evaluation of the current strategy by government. This issue begins to touch on the following conclusion.

**Current broadband strategies based on 4G or 5G generally overlook temporal evolution**. This is to the detriment of achieving long-term objectives. For example, the UK's telecom regulator Ofcom focuses on three-year cycles to assess the mobile market [79], meaning there is a short-term perspective on the decisions for the various broadband strategies employed. Our conjecture, informed by the findings of this analysis, is that this type of short-term horizon is too limited. Thus, there needs to be greater appreciation for how cellular infrastructure will be upgraded as each generation is deployed, for example, from 4G to 5G to 6G. This is not to say governments should attempt to predict or forecast the market or indeed technological development for telecom technologies. Instead, there should be greater recognition that telecom regulators can introduce infrastructure policies that encourage the deployment of favorable technologies which will provide long-term benefits. In the case of the assessment presented in this paper, an example would be developing supportive policies which encourage greater fiber deployment. Fiber in the ground that can be easily accessed by MNOs and other telecom providers will have long-term benefits. Indeed, those benefits are well documented, with society developing considerably when citizens have greater opportunities to use digital technologies. Moreover, the economy benefits from efficient infrastructure, in terms of greater productivity improvements, and how this contributes to growth in a nation's Gross Domestic Product (which in turn generates greater tax revenue). Universal broadband is fundamentally a good thing, but we need to consider the evolution over time between generations of technology.

**6G R&D efforts need to remember the other cost factors that will influence global broadband coverage**. In 5G, many new and fantastic ways to deliver higher capacity were introduced, and in turn, help to reduce the cost per bit of data transfer (e.g. 64x64 Massive MIMO). However, this is one example of a uniquely dense urban solution for providing capacity. In fact, 5G in general did very little to help deploy broadband for rural and hard-to-reach locations. Granted, some research groups did undertake efforts on this topic, but generally, it was a small-scale activity, focusing mainly on rural deployment. Thankfully, many have already recognized the limitations of 5G in this regard and have attempted to bring this up the agenda for 6G R&D and future standardization. This is no doubt highly important and the assessment carried out in this paper supports that approach while also wishing to contribute conclusions of our own. The challenge will be in helping to deploy wide-area connectivity solutions in low-APRU environments which are able to maximize efficiency in terms of spectrum and energy use, and therefore cost.

**There needs to be a greater emphasis on independent open-source cost assessment for candidate 6G technologies in earlier phases of standardization**. In many ways, the cost assessment of 5G technologies was very much an afterthought. Indeed, the majority of peer-reviewed papers on 5G cost assessment occurred very late in the standardization cycle from approximately 2018 onwards [87], [134]–[136]. This mistake must not be repeated, and without undertaking independent assessment of these technologies in advance, 6G will fall into the same position. Many of the standardized technologies were a set of very urban solutions, rather than the engineering community presenting technological options for a wide range of urban *and* rural connectivity problems. Moreover, the 5G standardization process lacked the use of open-source tools widely used across the software development community, but which would help identify the best technological candidates for standardization. More work should be openly published which evaluates the use of different network architectures in heterogenous deployment scenarios. This should provide compelling evidence for researchers to help support those technologies which provide the best solutions in terms of cost-efficiency.

Having identified four key conclusions, future research will now be discussed. Firstly, there needs to be more assessment evaluating the trade-off in cost for remote locations between 5G cellular and newly deployed Low Earth Orbit (LEO) satellite broadband constellations, such as those being launched by Space X (Starlink), OneWeb and Blue Origin (Kuiper). Given the latency provided by LEO broadband satellites is now highly competitive with terrestrial options, it may be more affordable to use this connectivity to provide small, single villages with connections where the number of residents is under the viable threshold for cellular technologies to be deployed. Secondly, there also needs to be more assessment evaluating the size of the benefits from enhanced digital connectivity because this would help more robust cost-benefit assessment in government be undertaken in relation to the provision of reliable broadband connectivity.

This paper contributes to the literature in three specific ways. Firstly, in assessing how different 4G and 5G

strategies quantitatively perform in viably delivering universal broadband coverage. Secondly, in evaluating the impact that spectrum price changes have on coverage-focused universal broadband strategies. Finally, in identifying conclusions to inform current 5G policies and future 6G standardization and deployment.

## ACKNOWLEDGMENT

The authors would like to thank their respective institutions for funding support, as well as anonymous reviewers of the paper. Luis Mendo kindly provided comments on an earlier version of the paper.

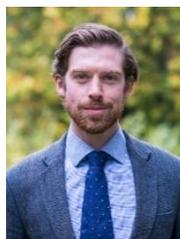
**EDWARD J. OUGHTON** received the M.Phil. and Ph.D. degrees from Clare College, at the University of Cambridge, U.K., in 2010 and 2015, respectively. He later held research positions at both Cambridge and Oxford. He is currently an Assistant Professor in the College of Science at George Mason University, Fairfax, VA, USA, developing open-source research software to analyze digital infrastructure deployment strategies. He received the Pacific Telecommunication Council Young Scholars Award in 2019, Best Paper Award 2019 from the Society of Risk Analysis, and the TPRC48 Charles Benton Early Career Scholar Award 2021.

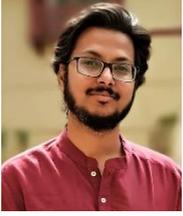

**ASHUTOSH JHA** completed his Ph.D. from the Indian Institute of Management, Calcutta in 2020, and his B.E. (Hons.) in Electrical and Electronics Engineering from BITS-Pilani, Goa, India, in 2008. He is currently an Assistant Professor in the Information Management Group at SPJIMR, Mumbai. His research interests include Adoption and Diffusion of Emerging Technologies, Economics of Next Generation Networks, and Development Informatics. He has published in journals such as Information Systems Frontiers, Technological Forecasting and Social Change and IIMB Management Review.